\documentclass[aps,pra,amsmath,amssymb,amsfonts,superscriptaddress,reprint,twocolumn]{revtex4-1}

\usepackage{dcolumn}    
\usepackage{graphicx}
\usepackage{amsmath, bbm}
\usepackage{latexsym}
\usepackage{amsfonts}   
\usepackage{amssymb}
\usepackage{array}      
\usepackage[dvipsnames]{xcolor} 
\usepackage{ulem}
\usepackage{soul}
\usepackage{dsfont}
\usepackage{physics}
\usepackage{bm}
\usepackage[colorlinks=true,linkcolor=blue,urlcolor=blue,citecolor=blue,pdfusetitle]{hyperref}
\usepackage{orcidlink}

\AtBeginDocument{%
    \newwrite\bibnotes
    \def\bibnotesext{Notes.bib}
    \immediate\openout\bibnotes=\jobname\bibnotesext
    \immediate\write\bibnotes{@CONTROL{REVTEX41Control}}
    \immediate\write\bibnotes{@CONTROL{%
    apsrev41Control,author="08",editor="1",pages="1",title="0",year="1"}}
     \if@filesw
     \immediate\write\@auxout{\string\citation{apsrev41Control}}%
    \fi
}%

\makeatletter
\newsavebox{\@brx}
\newcommand{\llangle}[1][]{\savebox{\@brx}{\(\m@th{#1\langle}\)}%
  \mathopen{\copy\@brx\kern-0.5\wd\@brx\usebox{\@brx}}}
\newcommand{\rrangle}[1][]{\savebox{\@brx}{\(\m@th{#1\rangle}\)}%
  \mathclose{\copy\@brx\kern-0.5\wd\@brx\usebox{\@brx}}}
\makeatother

\newcommand{\J}{\hat{J}}

\begin{document}

\title{
Attaining Quantum Sensing Enhancement from Monitored Dissipative Time Crystals
}
\author{Eoin O'Connor\,\orcidlink{0000-0002-8900-6649}}
\email{eoin.oconnor@unimi.it}
\affiliation{Dipartimento di Fisica {\it "Aldo Pontremoli"}, Universit\`a degli Studi di Milano, I-20133 Milano, Italia}
\author{Victor Montenegro\,\orcidlink{0000-0003-3846-3863}}
\affiliation{College of Computing and Mathematical Sciences, Department of Applied Mathematics and Sciences, Khalifa University of Science and Technology, 127788 Abu Dhabi, United Arab Emirates}
\affiliation{Institute of Fundamental and Frontier Sciences, University of Electronic Science and Technology of China, Chengdu 610051, China}
\affiliation{Key Laboratory of Quantum Physics and Photonic Quantum Information, Ministry of Education, University of Electronic Science and
Technology of China, Chengdu 611731, China}
\author{Francesco Albarelli\,\orcidlink{0000-0001-5775-168X}}
\affiliation{Scuola Normale Superiore, I-56126 Pisa, Italy}
\affiliation{Università di Parma, Dipartimento di Scienze Matematiche, Fisiche e Informatiche, I-43124 Parma, Italy}
\affiliation{INFN—Sezione di Milano-Bicocca, Gruppo Collegato di Parma, I-43124 Parma, Italy}
\author{Matteo G. A. Paris\,\orcidlink{0000-0001-7523-7289}}
\affiliation{Dipartimento di Fisica {\it "Aldo Pontremoli"}, Universit\`a degli Studi di Milano, I-20133 Milano, Italia}
\author{Abolfazl Bayat\,\orcidlink{0000-0003-3852-4558}}
\affiliation{Institute of Fundamental and Frontier Sciences, University of Electronic Science and Technology of China, Chengdu 610051, China}
\affiliation{Key Laboratory of Quantum Physics and Photonic Quantum Information, Ministry of Education,
University of Electronic Science and Technology of China, Chengdu 611731, China}
\affiliation{Shimmer Center, Tianfu Jiangxi Laboratory, Chengdu 641419, China}
\author{Marco G. Genoni\,\orcidlink{0000-0001-7270-4742}}
\email{marco.genoni@unimi.it}
\affiliation{Dipartimento di Fisica {\it "Aldo Pontremoli"}, Universit\`a degli Studi di Milano, I-20133 Milano, Italia}
\affiliation{Shimmer Center, Tianfu Jiangxi Laboratory, Chengdu 641419, China}
%
\begin{abstract}
This study investigates quantum-enhanced parameter estimation through continuous monitoring in open quantum systems that exhibit a dissipative time crystal phase. We first analytically derive the global quantum Fisher information (QFI) rate for boundary time crystals (BTCs), demonstrating that within the time-crystal phase, the ultimate precision exhibits a cubic scaling with the system size, $f_{\mathrm{global}}\sim N^3$. We then generalize this finding to a broader class of dynamics, including the transverse collective dephasing (TCD) model, which achieves a time-crystal phase through a closing Liouvillian gap without requiring a dissipative phase transition. We proceed to numerically demonstrate that this maximal global QFI rate is experimentally attainable for both the BTC and TCD models, even at finite system sizes, via continuous homodyne and photodetection. Moving towards practical implementations, we analyze the precision limits under inefficient detection, revealing a critical difference: for BTC dynamics, inefficiencies asymptotically restore a classical scaling, and only a constant-factor quantum advantage remains possible. In contrast, for TCD dynamics, a super-classical scaling is still in principle observable, and our numerical simulations confirm its presence, even under inefficient measurement conditions, establishing the TCD model as a highly robust platform for quantum metrology.
%
\end{abstract}
\date{\today}

\maketitle

\section{Introduction}
Quantum metrology aims to surpass classical precision limits in parameter estimation by leveraging uniquely quantum resources such as entanglement and squeezing~\cite{Giovannetti2011}.
A particularly powerful paradigm is critical quantum metrology, which exploits the enhanced sensitivity near phase transitions in many-body quantum systems—either in the ground or thermal states of critical Hamiltonians~\cite{Zanardi2008, Invernizzi2008,Salvatori2014a,Rams2018,Sarkar2022topological,Gietka2022,DiFresco2022,Hotter2024,Yu2024a,Mihailescu2024,Ostermann2024,Montenegro2025b} or in the steady states of systems undergoing Floquet~\cite{Mishra2021driving,Mishra2022integrable,shukla2025prethermal} or dissipative~\cite{Fernandez2017,ivanov2020enhanced,DiCandia2023,Alushi2024,Beaulieu2025a,Gorecki2025a,liu2026enhanced} phase transitions.
Criticality-based quantum sensors have also been experimentally realized in various physical platforms, including Rydberg atoms~\cite{Ding2022enhanced,wang2026quantum,liu2026enhanced}, solid state systems~\cite{liu2021experimental,Wu2024experimental,moon2026sensing}, photonic setups~\cite{xiao2025observation} and superconducting devices~\cite{yu2025experimental,Beaulieu2025criticalityEnhanced,li2025nonequilibrium}. 

Boundary time crystals are among the prominent systems that exhibit dissipative quantum phase transitions due to competition between coherent Rabi oscillations and collective damping~\cite{Agarwal1977,Carmichael1977,puri1979,Drummond1980,Morrison2008,ieminiBoundaryTimeCrystals2018}. 
Recently, they have received great attention from their fundamental and practical properties~\cite{montenegroQuantumMetrologyBoundary2023,CabotPRA2023,IeminiPRA2024,cabotContinuousSensingParameter2024,Gribben2025,paulino2026thermodynamics,passarelli2025nonstabilizerness,Cabot2025,midha2025arXiv,lee2025arXiv,mattes2025designing,das2026stabilizing} and they have been experimentally investigated in free-space atomic ensembles~\cite{Ferioli2023}.

Quantum-enhanced sensing in such systems was initially evidenced at the phase transition between the static and the time crystalline phase, where the quantum Fisher information (QFI) of the steady state scales super-linearly $F_Q \sim N^{4/3}$, with $N$ being the system size~\cite{montenegroQuantumMetrologyBoundary2023}.
Notice that this and the other scalings in $N$ are valid by assuming a dissipation rate that scales as $1/N$ (the so-called Kac rescaling), as needed for a well-defined thermodynamic limit~\cite{Benatti2016,Benatti_2018,Carollo_2024,kokalj2025Kac}.
However, this enhancement vanishes, reverting to the standard quantum limit (SQL), when the probe preparation time $T$ is included in a full resource analysis, i.e. $F_Q/T \sim N$~\cite{montenegroQuantumMetrologyBoundary2023}. Indeed, the practicality of criticality-based sensors, including boundary time crystals, is constrained by several challenges: (i) resource-demanding probe preparation, which can erode quantum-enhanced precision; (ii) sophisticated measurement schemes that are often dependent on the unknown parameter; and (iii) a limited range around the critical point for achieving quantum-enhanced precision. The development of novel strategies to mitigate these constraints is highly desirable.

To address these challenges, a powerful approach is to extract information by continuously monitoring the environment in dissipative quantum systems~\cite{WisemanMilburn,AlbarelliPLA2024,Nurdin2022}. Such strategies have been widely studied in order to derive the corresponding estimation bounds~\cite{gammelmark2013a,gammelmarkFisherInformationQuantum2014,Genoni2017,DayouPRX,Khan2025,Yang2025b} or sense an external parameter in the evolution of the system~\cite{Ralph2017,Cortez2017,KiilerichPC,Fallani2022,Geremia2003,Madsen2004,Albarelli2017a,AmorosBinefa2021,amoros2025noisy,Albarelli2018restoringheisenberg,albarelli2020quantum,Rossi2020PRL}. 
Recently, attention has turned to combining continuous monitoring with many-body systems undergoing dissipative phase transitions~\cite{Ilias2022,cabotContinuousSensingParameter2024,lee2025arXiv,midha2025arXiv,mattes2025designing,Cabot2025}.
Notably, by exploiting the continuous monitoring of the environment in boundary time crystals, the quantum enhanced precision is restored and even significantly improves to $F_Q/T \sim N^3$~\cite{cabotContinuousSensingParameter2024,midha2025arXiv,lee2025arXiv}. Several open questions arise, including: 
i) Is this quantum enhancement specific to BTCs or can it be generalized to a wider family of dissipative evolutions leading to a time-crystal phase at steady-state~\cite{kesslerEmergentLimitCycles2019, Lledo_2020, seiboldDissipativeTimeCrystal2020, piccitto2021symmetries, souzaSufficientConditionGapless2023, nemeth2025solving}? ii) What is the role of the dissipative phase transition in observing this quantum enhancement? iii) Can one achieve such quantum enhancement over a wide range of parameters by using a simple measurement scheme, such as homodyne detection~\cite{orenes2022improving,duan2025concurrent}? iv) How sensitive is such enhancement to imperfections in the measurement setup?

Here, we answer the above questions by identifying a broad family of collective spin models that yield the same Fisher information scaling as the BTC. Interestingly, this family includes the transverse collective dephasing (TCD) evolution, where a quantum non-demolition (QND)-like dissipative term is engineered, typically via a dispersive interaction with a cavity mode.
This has been widely studied both from a theoretical
~\cite{doherty1999feedback,thomsen2002spin,Geremia2003,barberena2024trade,Madsen2004,MolmerMadsen2004,Albarelli2017a,Rossi2020PRL,AmorosBinefa2021,caprotti2024analysis} and experimental~\cite{kuzmich2000generation,appel2009mesoscopic,schleier2010states,bohnet2014reduced,cox2016deterministic,hosten2016measurement,serafin2021nuclear,orenes2022improving,huang2023observing,duan2025concurrent} point of view.
For all these models, we demonstrate that the enhancement arises from the closure of the Liouvillian gap induced by the competition between driving and dissipation, rather than the criticality of a phase transition.
Furthermore, we derive a general upper bound for the QFI rate based on the spectral properties of the Liouvillian and the fluctuations of the generator of the parameter encoding, and show how this bound correctly reproduces the scalings derived for our general collective spin models.
In particular, we show that when the dissipation rate is constant, the QFI rate shows a Heisenberg-limited $N^2$ scaling.
However, considering the thermodynamic-limit rescaling of the dissipation rate $\kappa \to \kappa/N$, one recovers the enhanced scaling $N^3$, already mentioned above.

Remarkably, we then show that for the BTC and the TCD model, both continuous homodyne and continuous photodetection attain the promised quantum-enhanced scaling, saturating the QFI rate. Another key result of our analysis is to consider finite efficiency in the continuous monitoring, observing how a notable difference between the BTC and the TCD model appears in this scenario. Using tools from noisy quantum metrology~\cite{demkowicz2017adaptive,Wan2022,Kurdzialek2023a,Das2025a}, we derive the fundamental precision bound for detection efficiency $\eta<1$ for the BTC model, showing that inefficiencies asymptotically restore SQL scaling, permitting only a constant-factor quantum advantage, scaling as $(1-\eta)^{-1}$. Remarkably, in the TCD model, such a no-go theorem does not hold, and we numerically show that a super-classical (efficiency-dependent) scaling can still be observed also in the presence of measurement imperfections.
\\

%
%
\section{Results}
\subsection{Quantum estimation via continuous monitoring}
\label{sec:QETmonitoring}
The precision of estimating a parameter, $\theta$, encoded in a quantum state $\rho_{\theta}$ 
and measured via a positive-operator valued measure (POVM) $\{\hat{\Pi}_x\}$ is limited by the (classical) Cram\'er-Rao bound. This bound relates the variance of an unbiased estimator $\widetilde{\theta}$ to the (classical) Fisher information (FI), $F_{\{\Pi_x\}}(\rho_\theta)$, via $\operatorname{Var}(\widetilde{\theta}) \ge 1/F_{\{\hat{\Pi}_x\}}(\rho_\theta)$~\cite{HelstromBook,CavesBraunstein,Matteo2009}. The FI is a simple function of the conditional probability distribution $p(x|\theta) = \hbox{Tr}[\rho_\theta \hat{\Pi}_x]$, via the formula $F_{\{\hat{\Pi}_x\}}(\rho_\theta) = \sum_x p(x|\theta) (\partial_\theta \log p(x|\theta))^2$. By maximizing it over all the possible POVMs, one obtains the quantum Fisher information (QFI), $F_Q(\rho_\theta)$, which depends only on the quantum state $\rho_\theta$ and sets the ultimate limit on the estimation precision via the quantum Cramér-Rao bound (QCRB), $\text{Var}(\widetilde{\theta}) \ge 1/F_Q(\rho_\theta)$~\cite{HelstromBook,CavesBraunstein,Matteo2009}.

In open quantum systems, continuous monitoring of the environment into which the system dissipates is a powerful and natural technique for parameter estimation. The total information that can be obtained from such protocols, which combines the information from the continuous measurement record with a final strong measurement on the system, is bounded by the global QFI of the combined system-environment state \cite{gammelmarkFisherInformationQuantum2014}. We consider physical scenarios where the parameter $\theta$ is encoded solely in the system Hamiltonian $\hat{H}_\theta$, and the system's unconditional evolution is described by a Lindblad master equation,
$\frac{d \rho_\theta}{d t} = \mathcal{L}_\theta(\rho_\theta) = -i[\hat{H}_\theta, \rho_\theta] + \sum_k \mathcal{D}[\hat{c}](\rho_\theta)$
where $\mathcal{D}[\hat{L}](\rho) = \hat{L}\rho \hat{L}^\dagger - \frac{1}{2}\{ \hat{L}^\dagger \hat{L}, \rho\}$ is the standard Lindblad dissipator. In this case the global QFI accumulated up to time $T$ can be expressed as a two-time correlation function of the parameter's generator~\cite{gammelmarkFisherInformationQuantum2014,Ilias2022}
\begin{equation}
    F_{\text{global}}(T) = 2 \int_0^T \dd\tau \int_0^T \dd\tau' \, \langle\{\delta \hat{O}(\tau'), \delta \hat{O}(\tau)\}\rangle.
    \label{eq:global_qfi_hamiltonian}
\end{equation}
In this expression, $\hat{O} := \partial_\theta \hat{H}_\theta$ is the Hermitian operator that generates shifts in the parameter $\theta$.
The Heisenberg picture gives the time evolution of the operator for open systems, $\hat{O}(\tau) = e^{\mathcal{L}_\theta^\dagger \tau}(\hat{O})$, where $\mathcal{L}_\theta^\dagger$ is the adjoint of the Lindbladian superoperator. The term $\delta \hat{O}(\tau) = \hat{O}(\tau) - \langle \hat{O}(\tau) \rangle$ represents the fluctuation of the time evolved operator, where the expectation value $\langle\cdot\rangle = \Tr[\cdot \rho(0)]$ is taken with respect to the system's initial state $\rho(0)$.

The continuous monitoring of the environmental degrees of freedom causes a conditional evolution of the quantum system.
Different measurement strategies lead to different {\it unravellings} of the master equation,
that is, to different stochastic master equations for the conditional states $\rho_\theta^{(c)}$~\cite{AlbarelliPLA2024}. The most paradigmatic examples and experimentally relevant unravellings correspond to continuous photodetection and homodyne detection, leading respectively to a quantum-jump-like and a diffusive evolution for the conditional states (see Methods~\ref{a:SME} for more details on such stochastic master equations).

The total information on a parameter $\theta$ that can be extracted in such experiments depends on the specific chosen measurement. It is quantified by the {\it unravelling} FI $F_{\text{unr}}(T)$~\cite{Albarelli2017a,Albarelli2018restoringheisenberg}, which is composed of the classical FI from the continuous measurement signal, $F_{\text{signal}}(T)$~\cite{gammelmark2013a}, and the QFI of the conditional states $\rho_\theta^{(c)}$, averaged over all measurement records, $\mathbb{E}[F_Q(\rho_\theta^{(c)}(T))]$. This sum is fundamentally bounded by the global QFI in Eq.~\eqref{eq:global_qfi_hamiltonian}:
\begin{equation}
    F_{\text{unr}}(T) = F_{\text{signal}}(T) + \mathbb{E}\left[F_Q(\rho_\theta^{(c)}(T))\right] \le F_{\text{global}}(T).
    \label{eq:info_decomposition_theta}
\end{equation}
In the long-time limit, $T \to \infty$, assuming the system reaches a unique steady state, both $F_{\text{global}}$ and $F_{\text{signal}}$ typically scale linearly with $T$~\cite{gammelmarkFisherInformationQuantum2014,Ilias2022}.
It is therefore convenient to analyze and compare the signal FI rate, $f_{\text{signal}} = \lim_{T\to\infty} \mathcal{F_{\text{signal}}}(T)/T$ with the corresponding global QFI rate $f_{\text{global}}$, as the contribution from the final strong measurement $\mathbb{E}[F_Q(\rho_\theta^{(c)}(T))]/T$ becomes negligible in this regime.

However, there is no guarantee that $f_{\text{signal}} = f_{\text{global}}$, even with perfect monitoring efficiency.
The inequality in Eq.~\eqref{eq:info_decomposition_theta} can in fact be strict: this occurs when the specific measurement performed on the environment fails to capture all the information that leaks from the system. Whether saturation is possible, i.e., whether $f_{\text{signal}} = f_{\text{global}}$, depends critically on the interplay between the Hamiltonian dynamics, the dissipative channels, and the chosen measurement scheme~\cite{mattes2025designing}.

\subsection{Boundary time crystal}
\label{sec:btc}
The system we first analyze is a paradigmatic model for dissipative phase transitions and time-crystals \cite{Agarwal1977,Carmichael1977,Morrison2008,ieminiBoundaryTimeCrystals2018}. It consists of $N$ non-interacting spins, whose collective behavior is conveniently described by the total angular momentum operators $\J_\alpha = \frac{1}{2}\sum_{j=1}^N \hat{\sigma}_\alpha^{(j)}$ for $\alpha \in \{x,y,z\}$, where $\hat{\sigma}_\alpha^{(j)}$ are the Pauli matrices for the $j$-th spin. The corresponding raising and lowering operators are $\J_{\pm} = \J_x \pm i \J_y$.

The system's dynamics are governed by a Lindblad master equation, which features a competition between coherent driving, described by a  Hamiltonian, $H = \omega \J_x$, and collective decay described by a single dissipative channel:
\begin{equation}
    \frac{d\rho}{dt} = \mathcal{L}(\rho) = -i\omega[\J_x, \rho] + \frac{2\kappa}{N}\mathcal{D}[\J_-]\rho.
    \label{eq:btc_master_equation}
\end{equation}
We label this the boundary time crystal (BTC) model. The jump operator $\J_-$ induces transitions that lower the total spin projection $\J_z$. The dissipation rate is scaled by $2/N$, which ensures a well-defined thermodynamic limit ($N\to\infty$) where the dynamics become independent of the system size~\cite{Benatti2016,Benatti_2018,Carollo_2024}.

In the large-$N$ limit, the system exhibits a dissipative phase transition at a critical point $\omega_c = \kappa$. 
When dissipation is strong relative to the driving ($\omega < \kappa$), the system is pulled towards a unique, non-oscillating steady state characterized by a large negative polarization, $\langle \J_z \rangle < 0$. Conversely, when the driving is sufficiently strong ($\omega > \kappa$), it prevents the system from fully relaxing into a static configuration. The system instead enters a dynamic phase of persistent, self-sustained oscillations. In this regime, the expectation values of spin observables, such as $\langle \J_y(t) \rangle$ and $\langle \J_z(t) \rangle$, exhibit robust oscillations with a frequency determined by the system parameters.
These critical properties allow the system to surpass the standard quantum limit for estimation of the system driving frequency $\omega$: at the critical point, the QFI of the steady state scales indeed as $F_Q(\rho_{\mathrm{ss}}) \sim N^{4/3}$~\cite{montenegroQuantumMetrologyBoundary2023}. However, if the time $T_{\mathrm{ss}}$ needed to reach the steady state is accounted for as a resource, the enhancement vanishes, and the QFI rate reverts to the SQL scaling, i.e. $f_Q(\rho_{\mathrm{ss}}) =F_Q(\rho_{\mathrm{ss}})/T_{\mathrm{ss}}  \sim N$.

\subsection{Global QFI rate Scaling for the BTC model}
\label{sec:qfi_rate_scaling}
\begin{figure}[t]
\begin{center}
\includegraphics[angle=0,width=0.98\linewidth]{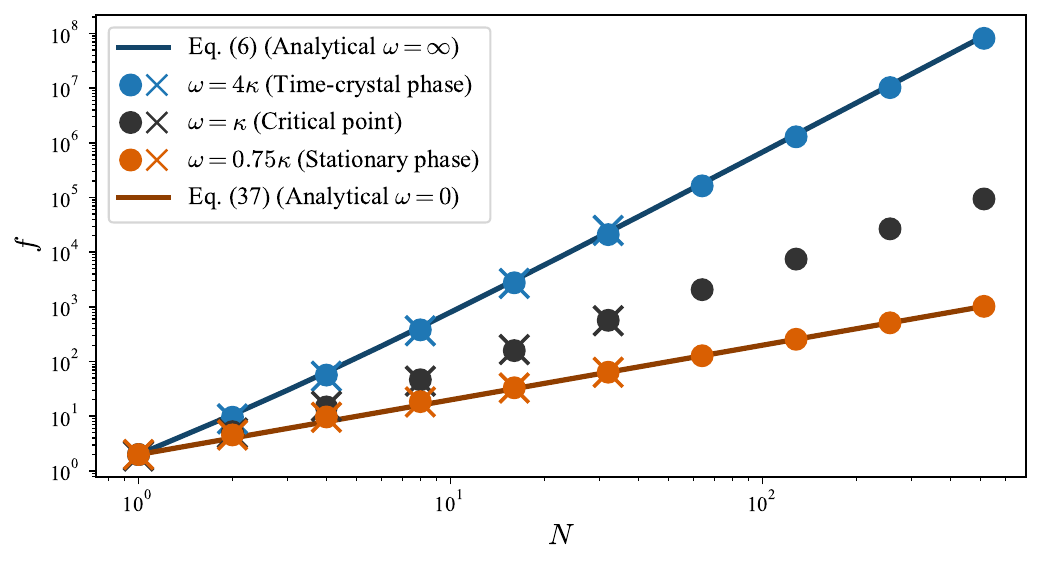} 
\end{center}
\caption{The steady-state global QFI rate $f_{\text{global}}$ at $\omega = 4\kappa$ (blue circles), $\omega = \kappa$ (black circles) and $\omega = 0.75\kappa$ (orange circles). We demonstrate that the steady-state Fisher information rate from homodyne detection $f_{\text{signal}}$ is identical to $f_{\text{global}}$ at all three values $\omega = 4\kappa$ (blue crosses), $\omega = \kappa$ (black crosses), and $\omega = 0.75 \kappa$ (orange crosses). Additionally, we show that our analytic results for $f_{\text{global}}$ in the extreme time-crystal (Eq.~\eqref{eq:fglobal_analytic}, navy line) and static limit (Eq.~\eqref{eq:static_limit}, orange line) closely match the numerical results with $\omega = 4\kappa$ and $\omega = 0.75\kappa$, respectively.}
\label{f:fig1}
\end{figure}

We now determine the ultimate precision bound for frequency estimation in the BTC model by calculating the global QFI defined in Eq.~\eqref{eq:global_qfi_hamiltonian}. Although Eq.~\eqref{eq:global_qfi_hamiltonian} cannot be solved in full generality, it simplifies whenever $e^{\mathcal{L}^\dagger t}[\hat O]$ is analytically tractable. In the Schr\"odinger picture, the global QFI is given by,
\begin{equation}
    F_{\text{global}}(T) \!=\! 4 \int_0^T d\tau \int_0^\tau d\tau' \, \text{Tr}\!\left[ \delta \hat{O} e^{\mathcal{L}(\tau-\tau')} [\{\delta \hat{O}, \rho(\tau')\}] \right].
    \label{eq:Fglobal}
\end{equation}
Using the relation $\text{Tr}\!\left[ \hat{A}^\dag e^{\mathcal{L}t} [\hat{B}] \right] = \text{Tr}\!\left[ (e^{\mathcal{L}^\dag t}[\hat{A}])^\dag  \hat{B} \right]$ we can instead evaluate the Heisenberg picture evolution of $\delta \hat{O}$.
It was recently shown that in the extreme time-crystal limit ($\omega/\kappa \to \infty$), it is possible to diagonalize the BTC Lindbladian in terms of the superspin operators $\hat{S}^2$ and $\hat{S}_x$~\cite{nemeth2025solving}. In Methods~\ref{a:Lindblad_diagonalization} we demonstrate that $\J_x$ is a right eigenoperator of $\mathcal{L}$ and $\mathcal{L}^\dag$ with eigenvalue $-\kappa/N$. For simplicity, we will choose the optimal initial condition, $\langle \J_x \rangle_0 = 0$. Then the integrand takes the form
\begin{equation}
    \text{Tr}\!\left[ \J_x e^{\mathcal{L}(\tau-\tau')} [\{\J_x, \rho(\tau')\}] \right] = e^{-(\tau-\tau')\kappa/N} \langle \hat{J}_x^2 \rangle_{\rho(\tau')}
    \label{eq:transient_integrand}
\end{equation}
In the long-time limit, the system approaches its steady state, which in this regime is the maximally mixed state~\cite{puri1979,Drummond1980}. The global QFI rate can then be evaluated analytically:
\begin{align}
    f_{\text{global}} &= 8 \int_0^\infty d\tau \, \frac{e^{-\tau\kappa/N}}{N+1} \text{Tr}[\hat{J}_x^2] \nonumber \\
    &= \frac{2 N^2(N+2)}{3\kappa}.
    \label{eq:fglobal_analytic}
\end{align}
This is the first main result of our work; it rigorously quantifies a non-trivial $N^3$ scaling with the system size.
As shown in Fig.~\ref{f:fig1}, this analytical result provides excellent agreement with numerical evaluation of Eq.~\eqref{eq:global_qfi_hamiltonian}, even though the latter was obtained at $\omega=4\kappa$ rather than the extreme time-crystal limit. This further confirms some recent results that were also suggesting a $N^3$ scaling in the time crystal phase~\cite{cabotContinuousSensingParameter2024,midha2025arXiv,lee2025arXiv} (we refer to Methods~\ref{a:thermodynamiclimit_scaling} for a discussion about the different scaling that one obtains if one does not consider the proper {\it thermodynamic-limit induced rescaling} of the collective damping parameter in the master equation~\eqref{eq:btc_master_equation}).

We can further analyze the transient dynamics by determining the time evolution of the second moment.
From the master equation~\eqref{eq:btc_master_equation} one finds that the second moment of $\hat{J}_x$ obeys
\begin{align}
    \frac{d\langle \J_x^2 \rangle_t}{dt} = -\frac{\kappa}{N}(3 \langle \J_x^2 \rangle_t - \langle \J^2 \rangle_t) \nonumber\\
    =  -\frac{\kappa}{N}\left(3 \langle \J_x^2 \rangle_t - \frac{N(N+2)}{4}\right).
\end{align}
Solving this ODE, we find
\begin{equation}
    \langle \hat{J}_x^2 \rangle_t= \frac{N(N+2)}{12} + \left(\langle \hat{J}_x^2 \rangle_0 - \frac{N(N+2)}{12}\right)e^{-3\kappa \tau'/N}.
\end{equation}
This allows us to analyse the transient regime of the global QFI: when $\kappa T \sim 1$, $F_{\text{global}}(T) \sim \langle \hat{J}_x^2 \rangle_0 T^2$, this tells us that the properties of the initial state play a role at early times, where a quadratic scaling in time $T^2$ can still be observed. When $\kappa T \sim N$, one obtains $F_{\text{global}}(T) \sim N^3 T$ independent of the initial state. In the long time limit, the initial state is not important; however, the time we must wait to reach the $N^3$ regime scales linearly with $N$ (notice that this is crucial so that the $N^2 T^2$ Heisenberg limit still applies at any time and for any value of $N$~\cite{Ilias2022}).
We also find that $F_{\text{global}}(T)/T$ is monotonically increasing in time towards its long-time limit~\eqref{eq:fglobal_analytic}; therefore, it is always beneficial to gather data for a long time on a single trajectory rather than resetting the system and repeating the experiment.

At criticality ($\omega=\kappa$), a finite-size scaling analysis indicates $f_{\text{global}}\sim N^{\zeta}$ with $\zeta \approx 5/3$ (see Supplementary Note 1 for more details on the derivation of this exponent), showing that the monitored time-crystal phase outperforms the sensing capabilities at the critical point in the long-time limit, as depicted in Fig.~\ref{f:fig1} and previously reported in~\cite{cabotContinuousSensingParameter2024}.

In the stationary phase ($\omega < \kappa)$ precision is reduced even further with only linear SQL scaling present. Our numerical results are backed up by an analytic derivation of this linear scaling in the static $\omega \to 0$ limit, $f_{\text{global}} = 2N/\kappa$; proof is provided in Sec.~\ref{a:static_limit}. In Fig~\ref{f:fig1} we plot numerical results for $f_{\text{global}}$ and $f_{\text{signal}}$ at $\omega = 0.75\kappa$ and show that they both converge to the $\omega = 0$ results in the large $N$ limit.

\subsection{Global QFI rate scaling for general dissipative time crystal models}
\label{sec:enhancement_generality}

The fact that the optimal metrological performance appears deep in the oscillatory phase naturally raises the question: is the dissipative phase transition essential for the observed $N^3$ scaling? To investigate this, we analyze a generalized model of collective spins, which reveals that the crucial ingredients are the geometry of the driving and dissipation terms, and the closure of the Liouvillian gap.

To clarify the underlying mechanism, we recast the dynamics in a basis-independent form. For any real vector $\vec a$, define the collective spin operator along $\vec a$ as
$\J_{\vec a} := \vec a \cdot \hat{\vec J}$, in this form the commutator satisfies
\begin{equation}
[\hat J_{\vec a},\hat J_{\vec b}] = i \hat J_{\vec a \times \vec b}.
\end{equation}
Consider a general Lindbladian for $N$ spins with a coherent driving in a direction $\vec{h}$ and collective dephasing along the axes $\vec{\ell}_i$. Since all jump operators are Hermitian, the Lindbladian takes the form
\begin{equation}
\mathcal{L}[\bullet]
= -i \omega [\hat J_{\vec h},\bullet]
- \frac{1}{2}\sum_i [\hat J_{\vec \ell_i},[\hat J_{\vec \ell_i},\bullet]].
\label{eq:generic_Lindblad}
\end{equation}
The generator for estimating the driving frequency $\omega$ is $\hat{O} = \J_{\vec h}$. Using the commutation relation above, the action on the generator is
\begin{equation}
\mathcal{L}[\hat J_{\vec h}] = \mathcal{L}^\dag[\hat J_{\vec h}]
= \frac{1}{2}\sum_i \hat J_{\vec \ell_i \times (\vec \ell_i \times \vec h)}.
\end{equation}
When a dissipative axis is transverse to the drive, i.e. $\vec \ell_i \cdot \vec h = 0$, the double cross product becomes
$\vec \ell_i \times (\vec \ell_i \times \vec h)= -\|\vec \ell_i\|^2 \vec h$. When a dissipative axis is parallel, the double cross product is 0. Assuming each dissipative axis is either parallel or perpendicular
\begin{equation}
    \mathcal{L}[\hat J_{\vec h}]
    = -\frac{1}{2}\left(\sum_{i,\vec \ell_i \cdot \vec h = 0}  \|\vec \ell_i\|^2\right)\hat J_{\vec h} := \kappa_{\text{eff}} \hat J_{\vec h}.
    \label{eq:general_eigenoperator}
\end{equation}
That is, $\hat J_{\vec h}$ is an eigenoperator of the Lindbladian (and of its adjoint) with eigenvalue $\kappa_{\text{eff}}$. In this transverse driving setup, the unique steady state is the maximally mixed state, and the two–time correlator entering the global QFI decays exponentially with that rate e. This allows us to generalize Eq.~\eqref{eq:transient_integrand} as
\begin{equation}
    \text{Tr}\left[ \hat{J}_{\vec h} e^{\mathcal{L}(\tau-\tau')} [\hat{J}_{\vec h} \rho(\tau')] \right] = e^{-(\tau-\tau')\kappa_{\text{eff}}} \langle \hat{J}_{\vec h}^2 \rangle_{\rho(\tau')},
\end{equation}
and Eq.~\eqref{eq:fglobal_analytic} as
\begin{equation}
    f_{\text{global}} = 8\int_0^\infty d\tau \, e^{-\kappa_{\text{eff}} \tau} \frac{\text{Tr}[J_{\vec h}^2]}{(N+1)} =  \frac{2N(N+2)}{3 \kappa_{\text{eff}}}.
    \label{eq:f_global_generic}
\end{equation}
This is our second main result: it generalizes our first main result, Eq.~\eqref{eq:fglobal_analytic}, to a broader class of models and highlights the closing Liouvillian gap as the origin of the enhanced scaling. The QFI rate scales as $N^2/\kappa_{\text{eff}}$, meaning if the effective decay rate scales as $\kappa_{\text{eff}} \sim N^{-\beta}$ for $\beta > 0$, this leads to enhanced scaling of $f_{\text{global}} \sim N^{2+\beta}$. As was shown in Methods~\ref{a:Lindblad_diagonalization}, the extreme time-crystal limit of the BTC model is a special case of Eq.~\eqref{eq:general_eigenoperator} with $\kappa_{\text{eff}} = \kappa/N$. Notice that even without a vanishing decay rate, these systems exhibit Heisenberg scaling. An interesting property of these models is that $f_{\text{global}}$ is independent of $\omega$, and there is no dissipative phase transition. However, the existence of a time-crystal phase (vanishing gap) is necessary to surpass Heisenberg scaling. 

Deriving an analytic form of $f_{\text{global}}$ for generic dephasing vectors $\{\vec{\ell}_i\}$ is generally not possible. However, when $\omega \|\vec h \| \gg \sum_i \|\vec{\ell}_i\|^2$ we can use a rotating wave approximation to evaluate $f_{\text{global}}$ (see Methods~\ref{a:gen_dephasing} for more details). By splitting each dephasing vector up into the part parallel and perpendicular to $\vec h$, $\vec{\ell}_i = \vec{\ell}_{i,\|} + \vec{\ell}_{i,\perp}$ we find that $\J_{\vec h}$ is once again an eigenoperator of the Lindbladian with $\kappa_{\text{eff}} = \frac12\sum_i \|\vec{\ell}_{i,\perp}\|^2$. This regime naturally occurs whenever we have enhanced scaling, i.e. whenever $\kappa_{\text{eff}} \sim N^{-\beta}$.

The importance of the interplay between driving and dissipation in realizing this enhancement can be further highlighted by considering a different driving axis. If we replace the transverse dephasing with a longitudinal one, i.e. a single jump operator with $\vec{\ell}\,\, \| \, \vec{h}$. This system still exhibits persistent oscillations of transverse spin components~\cite{nemeth2025solving}. However, these oscillations are not present in the photocurrent. In this case, the generator of the transformation, $\hat{O} = \J_z$, commutes with the jump operator $\J_z$. Consequently, only a finite amount of information about the parameter $\omega$ is leaked into the environment, and the signal Fisher information rate is zero for any monitoring scheme in the long time limit. This underscores that a transverse drive, which does not commute with the jump operators describing the monitored environmental coupling, is essential for achieving the quantum enhancement. We note that it is possible to achieve $N^2T^2$ scaling in the total QFI by preparing the system in a GHZ state and performing a ``no-information'' measurement~\cite{Albarelli2018restoringheisenberg,albarelli2020quantum}. However, this result requires perfect detection efficiency; whenever $\eta < 1$, the total Fisher information rate tends to 0 in the long time limit~\cite{albarelli2020quantum}.

To test our hypothesis that the metrological enhancement comes from the geometric interplay of driving and dissipation rather than the dissipative phase transition, we introduce a minimal model: the transverse collective dephasing (TCD) model.
\begin{equation}
    \frac{d\rho}{dt} = \mathcal{L}(\rho) = -i\omega[\J_x, \rho] + \frac{2\kappa}{N}\mathcal{D}[\J_z]\rho,
    \label{eq:TD_master_equation}
\end{equation}
This model corresponds exactly to the general form in Eq.~\eqref{eq:generic_Lindblad} with $\vec{h} \| \hat{x}$, and a single dephasing vector $\ell \| \hat{z}$. Therefore, it possesses a vanishing effective decay rate $\kappa_{\text{eff}} = \kappa/N$, leading to the exact same steady-state global QFI rate derived in Eq.~\eqref{eq:f_global_generic}. Crucially, unlike the BTC model, the TCD model does not exhibit a dissipative phase transition; its dynamics are solvable and exhibit persistent oscillations in the thermodynamic limit for any $|\omega|>0$. This allows us to isolate the influence of the Liouvillian gap closing from critical effects by comparing the BTC and TCD models. Additionally, this model is experimentally realisable in present-day trapped ion systems, and it is a prime candidate for continuous monitoring of many-body systems. As we will show in the Sec.~\ref{sec:inefficient_detection}, not only is this model simpler and easier to realize experimentally, it is also more resilient to experimental imperfections.

\subsection{An upper-bound for $f_{\text{global}}$}
To elucidate the physical origin of the enhanced scaling, we derive an upper bound for $f_{\text{global}}$ expressed in terms of the spectral properties of the unconditional Lindblad master equation~\cite{minganti2018}. Assuming a unique steady state $\rho_{\text{ss}}$, the global QFI rate can be written as
\begin{align}
    f_{\text{global}} &= 4 \int_0^\infty d \tau \Tr \left(\delta \hat{O} \, e^{\mathcal{L} \tau} \left[\{\delta \hat{O}, \rho_{ss}\} \right]  \right) \nonumber \\
    &= 4 \int_0^\infty d \tau \Tr \left(e^{\mathcal{L}^\dag \tau}[\delta \hat{O}]  \{\delta \hat{O}, \rho_{ss}\} \right)
\end{align}
We introduce the Drazin inverse of the adjoint Lindbladian~\cite{LandiPRXQuantum2024}, defined in vectorized notation as
\begin{equation}
    {\mathcal{L}^\dag}^+ = -\int_0^{\infty} e^{\mathcal{L}^\dag \tau} ( \mathbf{1} - |\hat{\mathbb{I}} \rrangle\llangle \rho_{ss}|).
\end{equation}
This superoperator effectively inverts the generator $\mathcal{L}^\dag$ within the subspace of operators orthogonal to the steady state (i.e., those with zero expectation value). Since $\delta \hat{O}$ has zero expectation value by construction, the integral simplifies to a compact form involving the symmetric logarithmic derivative (SLD) inner product~\cite{hayashi2017quantum}, $\langle \hat{A}, \hat{B} \rangle_S = \frac12\Tr(\hat{A}^\dag \, \{\rho_{\text{ss}}, \hat{B}\})$:
\begin{align}
    f_{\text{global}} &= 4\Tr \left({\mathcal{L}^\dag}^+[\delta \hat{O}]  \{\delta \hat{O}, \rho_{ss}\} \right) \nonumber \\
    &= 8\langle {\mathcal{L}^\dag}^+[\delta \hat{O}] , \delta \hat{O} \rangle_S.
\end{align}
We start by applying the Cauchy-Schwarz inequality
\begin{equation}
    f_{\text{global}} \leq 8 \|{\mathcal{L}^\dag}^+[\delta \hat{O}]\|_S \|\delta \hat{O}\|_S.
\end{equation}
The first term can be further bounded by introducing the operator norm induced by the SLD inner product, defined as
\begin{equation}
    \|\mathcal{K}\|_{\text{op},S} = \max_{\hat{A} \neq 0} \frac{\|\mathcal{K}[\hat{A}]\|_S}{\|\hat{A}\|_S}.
\end{equation}
We arrive at an upper bound on the global QFI rate
\begin{align}
    f_{\text{global}} &\leq 8\|\delta \hat{O}\|_S^2 \|{\mathcal{L}^\dag}^+\|_{\text{op},S} \nonumber \\
    &= 8\text{Var}_{\rho_{ss}}(\partial_\theta H) \|{\mathcal{L}^\dag}^+\|_{\text{op},S}
\end{align}
This bound represents a key theoretical contribution of this work.
It splits the metrological scaling into two distinct contributions: the steady-state variance of the parameter generator and the spectral properties of the dissipative dynamics.
To gather some intuition, we highlight that
$4 \text{Var}_{\rho_{\text{ss}}}(\partial_\theta H)$ equals the QFI of a phase-estimation protocol which considers as input any purification of the 
the steady-state $\rho_{\text{ss}}$ and a unitary encoding $U = e^{-i \phi (\partial_\theta H)}$ acting only on the system Hilbert space.
For local Hamiltonians, such as the BTC model, 
this term 
scales at most as $N^2$ (Heisenberg scaling). Consequently, any super-Heisenberg scaling (e.g., $N^3$) must arise from the divergent behavior of $\|{\mathcal{L}^\dag}^+\|_{\text{op},S}$.
This divergence can stem from a closing Liouvillian gap, or potentially from non-normality effects, i.e. arising when Lindbladian does not commute with its adjoint and thus the eigenoperators need not be orthogonal, such as exceptional points~\cite{minganti2019} or the skin effect~\cite{haga2021}.

To calculate $\|{\mathcal{L}^\dag}^+\|_{\text{op},S}$ we must map back to the standard Hilbert-Schmidt (HS) norm. In the vectorized representation, the steady-state anticommutator takes the form
\begin{equation}
    S = \frac{1}{2}\left(\hat{\mathbb{I}} \otimes \rho_{ss} + \rho_{ss}^{\intercal} \otimes \hat{\mathbb{I}}\right).
\end{equation}
The transformation $|\tilde{v}\rrangle = S^{\frac12} |v\rrangle$ recovers the standard HS norm. Therefore, the operator norm takes the form
\begin{equation}
    \|{\mathcal{L}^\dag}^+\|_{\text{op},S} = \|S^{\frac12}{\mathcal{L}^\dag}^+S^{-\frac12}\|_{\text{op}}.
\end{equation}
This can be evaluated by finding the smallest non-zero singular value of $S^{\frac12}\mathcal{L}^\dag S^{-\frac12}$.
While computing singular values for dense many-body Liouvillians is numerically demanding, this result rigorously identifies the mechanism of enhancement: a vanishing Lindbladian singular value \textit{within the metric defined by the steady state} that enables beyond Heisenberg scaling.

In the extreme time-crystal limit of the BTC model, the Lindbladian becomes normal, and the steady state approaches the maximally mixed state.
In this specific regime, $\|{\mathcal{L}^\dag}^+\|_{\text{op},S}$ is exactly the inverse of the Liouvillian gap ($\sim N/\kappa$), and the bound is saturated.
Similarly, the TCD model saturates this bound in the large-$N$ limit, where the Lindbladian superoperator becomes normal.

\begin{figure*}[t]
\begin{center}
\includegraphics[angle=0,width=\textwidth]{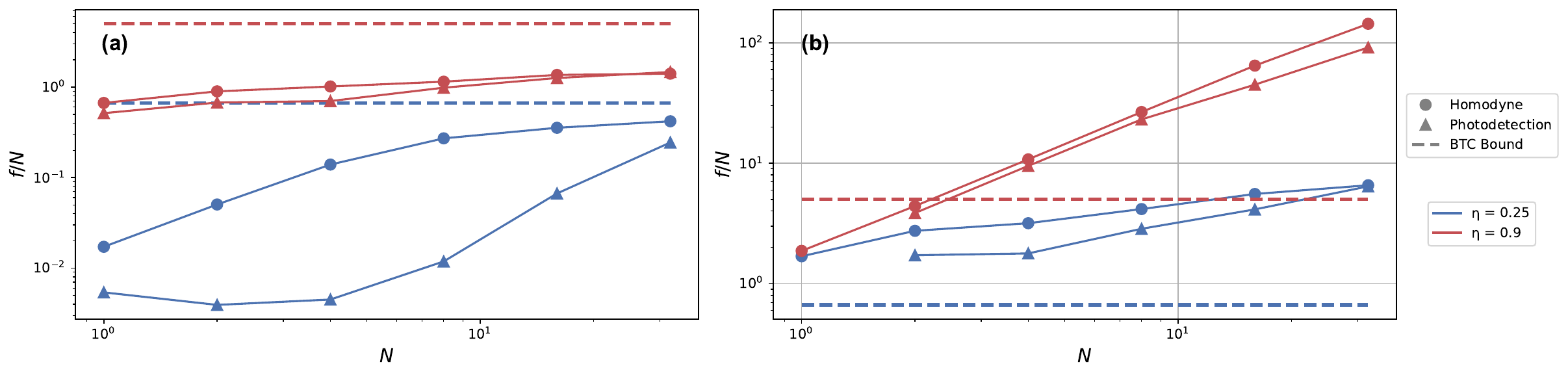} 
\end{center}
\caption{\textbf{(a)} The steady state Fisher information rate per spin of the measurement signal $f_{\text{signal}}/N$ for both homodyne (circles) and photodetection (triangles) in the BTC model. The rate is calculated for $\omega=4\kappa$ at two different measurement efficiencies $\eta = 0.9$ (red, top) and $\eta = 0.25$ (blue, bottom). The numerical results are compared against the theoretical bound from Eq.~\eqref{eq:inefficiencybound}.
\textbf{(b)} The same as \textbf{(a)} for the TCD model. The rate is calculated for $\omega=0.1\kappa$. We leave the bound from \textbf{(a)} for reference.
}
\label{f:fig2}
\end{figure*}

\subsection{Attaining the bound via continuous homodyne and photodetection}
\label{sec:HD_PD}
To check whether the ultimate bound can be attained by simple and practical monitoring strategies, we have considered the two most paradigmatic unravellings, corresponding to continuous photodetection and continuous homodyne detection~\cite{AlbarelliPLA2024}. 

We start by focusing on the BTC model: as regards photo-detection, the output signal is described at each time by a Poissonian increment $dN_t$ with average value $\mathbbm{E}[dN_t] = (2\kappa \hbox{Tr}[\rho^{(c)} \J_+\J_-] dt)/N$, while for homodyne is described by a continuous photocurrent $dy_t=2 \sqrt{2\kappa/N}\hbox{Tr}[\rho^{(c)} \J_y]\,dt + dw_t$ where $dw_t$ denotes a Wiener increment. Crucially, we have chosen the homodyne phase $\phi=\pi/2$. This ensures we monitor the persistent oscillations present in the $J_y$ operator; more details can be found in Methods~\ref{a:SME}).

By exploiting the techniques developed in~\cite{Albarelli2018restoringheisenberg}, we have evaluated the long time signal FI rate $f_{\textrm{signal}}$ for both strategies, by assuming a perfect detection efficiency $\eta=1$. As shown in Fig.~\ref{f:fig1}, homodyne detection attains the bound set by $f_{\textrm{global}}$ at the critical point and, more importantly, in the time-crystal phase ($\omega=4\kappa)$, thus achieving the cubic $N^3$ scaling even for finite $N$. We also find identical results for photodetection, confirming a recent proof in~\cite{mattes2025designing}. We also find that both strategies are saturating the corresponding bound also at the critical point ($\omega=\kappa$), where a lower scaling in $N$ is in fact observed. We remark that in the absence of monitoring, the steady-state QFI is much smaller for $\omega>\omega_c$ than it is at the critical point and also in the static phase ($\omega < \kappa$)~\cite{montenegroQuantumMetrologyBoundary2023}. This is because the steady-state in the extreme time-crystal limit can be approximated by the maximally mixed state~\cite{puri1979,Drummond1980}. These results show that continuous monitoring, purifying the system conditional states, allows more efficient information extraction about $\omega$ via the time-crystal oscillations.

We perform a similar analysis for the TCD model. In the case of photo-detection, the corresponding Poisson increment is characterized by its average value: $\mathbbm{E}[dN_t] = (2\kappa \hbox{Tr}[\rho^{(c)} \J_z^2] dt)/N$. For homodyne detection, it is important to choose the homodyne phase $\phi=0$, which allows us to monitor the oscillations in the $\J_z$ operator. Once again, both homodyne and photodetection extract all of the possible information, that is saturate the global QFI rate $f_{\text{global}}$. The numerical results are indistinguishable from the $\omega = 4\kappa$ data in Fig~\ref{f:fig1}. Concluding, for $\eta = 1$, the extreme time-crystal BTC and the TCD model perform identically at steady-state. As we will see, this behavior diverges significantly under more realistic constraints on the continuous monitoring performed on the system. 
%

\subsection{Inefficient detection}
\label{sec:inefficient_detection}
So far, our results assume ideal detection efficiency ($\eta=1$), which yields pure conditional states. Since this is experimentally unrealistic with current-day technology, it is crucial to analyze the precision bounds for $\eta<1$.

We start by focusing on the BTC dynamics. The dissipator in the unconditional master equation, Eq. \eqref{eq:btc_master_equation}, can be split as $(1-\eta)\mathcal{D}[\J_-] + \eta \mathcal{D}[\J_-]$.
The stochastic master equations corresponding to inefficient homodyne detection or photodetection are then
obtained by simply unravelling the second dissipator proportional to $\eta$~\cite{WisemanMilburn,AlbarelliPLA2024}.
As a consequence, the full evolution of the conditional BTC states obeys a stochastic differential equation of the form
\begin{equation}
\begin{split}
    {d\rho^{(c)}} =  -i & \omega  [\J_x, \rho^{(c)}]\,dt  \\
    & + \frac{2(1-\eta)\kappa}{N}\mathcal{D}[\J_-]\rho^{(c)}\,dt + \mathcal{S}(\rho^{(c)})\,
    \label{eq:btc_sme_efficiency}
    \end{split}
\end{equation}
where the last term $\mathcal{S}(\rho^{(c)})$ depends on the observed signal and describes the stochastic back-action of the continuous monitoring on the conditional states, and it satisfies $\mathbbm{E}[\mathcal{S}(\rho^{(c)})] = \frac{2\eta\kappa}{N} \mathcal{D}[\J_-]\rho^{(c)}$. 
This approach formally belongs to the general class of adaptive metrological strategies aided by auxiliary systems, in the presence of an unavoidable Markovian noise, represented here by the dissipator $\frac{2(1-\eta)\kappa}{N}\mathcal{D}[\J_-]$.
The ultimate precision bound that holds for all these protocols can be easily obtained just from the first two terms of
\eqref{eq:btc_sme_efficiency},
as described in \cite{demkowicz2017adaptive,Wan2022,Kurdzialek2023a,Das2025a} (see Methods~\ref{a:inefficient_bound}) obtaining
\begin{equation}
f_{\text{signal}} \leq \frac{N}{2(1-\eta)\kappa} \,.
\label{eq:inefficiencybound}
\end{equation}
Thus, for any non-unit detection efficiency, the precision reverts to the SQL, scaling linearly with N. This result holds irrespective of the driving frequency $\omega$.
However, we also observe that a constant factor enhancement $(1-\eta)^{-1}$ is possible in principle, 
and diverges in the limit of unit efficiency. This is consistent with the fact that for $\eta=1$ the scaling in $N$ is superlinear, as highlighted by our previous results.

Fig.~\ref{f:fig2}(a) shows the signal Fisher information rate per spin, $f_{\text{signal}}/N$, for the time-crystal phase ($\omega=4\kappa$) using both homodyne and photodetection at two different efficiencies. For both low ($\eta=0.25$) and high ($\eta=0.9$) efficiency, homodyne detection outperforms photodetection for smaller system sizes, though their performance appears to converge for larger $N$. In the low-efficiency regime, the data suggest that both strategies will saturate the ultimate bound as $N$ increases. Conversely, for high efficiency ($\eta=0.9$), while the rate increases monotonically with $N$, a persistent gap to the theoretical bound suggests it will not be reached, differing by a constant factor.
This result is understandable as one typically needs to implement non-trivial quantum error correction protocols to attain the bound~\cite{Zhou2019e}.
Tighter bounds for the continuous measurement with $\eta<1$ could in principle be obtained~\cite{Khan2025,Yang2025b}, but their numerical evaluation remains prohibitive for our model.

Shifting our attention to the TCD model described in Eq.~\eqref{eq:TD_master_equation} we find that the fundamental bound derived for the BTC model [Eq.~\eqref{eq:inefficiencybound}] does not apply. 
This is fundamentally because the Hamiltonian does not lie in the span of the Lindblad jump operators; in the language of Sec.~\ref{sec:enhancement_generality}, $\vec{h}$ cannot be written as a linear combination of $\{\vec{\ell}_i\}$.
For quantum estimation problems described by a noisy Lindblad dynamics of this kind, it is in principle possible to achieve a quadratic time scaling of the QFI by exploiting error correction techniques\cite{Zhou2018}.
More details on these fundamental precision bounds are provided in Methods~\ref{a:inefficient_bound}.
As this leaves open the possibility of quantum-enhanced scaling at finite efficiency, we investigate the $\eta < 1$ case via numerical simulations, plotting the results in Fig.~\ref{f:fig2}(b) for system size up to $N=32$.
In this parameter regime, we indeed observe that the signal FI rate $f_{\text{signal}}$ scales as $N^\alpha$ with $\alpha > 1$.
We find that $\alpha$ depends on $\eta$ and takes the values $\alpha = \{1.3,2.3\}$ for $\eta = \{0.25,0.9\}$ respectively.
This contrast constitutes our final main result: unlike the BTC, the TCD model is not bound by Eq.~\eqref{eq:inefficiencybound} and still present a super-classical scaling, despite their equivalence at $\eta = 1$.
The fact that the TCD model is not constrained by the finite efficiency bound means that it is a promising candidate for implementing error correction techniques that could restore $T^2$ scaling~\cite{demkowicz2017adaptive,Zhou2018} or increase $\alpha$ even if keeping a linear scaling in $T$~\cite{Gorecki2025a}.

%

\section{Discussion}
\label{sec:conclusions}
We have analytically derived the ultimate precision bound for frequency estimation in continuously monitored dissipative time crystals, showing a $N^3$ quantum-enhanced scaling in the time-crystal phase. 
We further demonstrated that, under ideal conditions ($\eta=1$), this bound can be attained in practice through continuous photodetection or homodyne detection, thus establishing experimentally accessible strategies capable of reaching the fundamental limit. 

We investigated the physical origin of this enhancement. By introducing the TCD model, we demonstrated that the dissipative phase transition is not a strict prerequisite. Instead, the enhancement emerges from the presence of strongly correlated oscillations in the photocurrent that become persistent in the thermodynamic limit. This insight decouples the metrological advantage from the complexity of critical many-body dynamics.

A crucial step towards realistic implementations is our analysis of finite detection efficiency. 
We derived the fundamental bound for the BTC model when $\eta<1$, showing that inefficiencies asymptotically restore SQL scaling. Still, a substantial constant-factor enhancement remains possible, diverging as $\eta\!\to\!1$. Numerical simulations indicate that continuous homodyne detection, in particular, approaches the ideal bound most closely at finite $N$.
In contrast, the TCD model is not subject to such a no-go theorem, due to the Hamiltonian not being in the span of the Lindblad jump operators.
Our numerical simulations show that it surpasses the standard quantum limit even when $\eta<1$, with an efficiency-dependent scaling.
This identifies the TCD model as a more robust candidate for an experimental verification of quantum enhanced scaling induced by a dissipative time crystal phase. 
%

While this work establishes the general features of quantum-enhanced sensing with monitored dissipative time crystals, several avenues for future research remain.
First of all, an experimental demonstration of such a quantum enhancement in the long-time stationary regime of continuous monitoring is still missing.
As mentioned above, the TCD model is closely related to QND interactions between atomic ensembles and light.
A prototypical realization of this setting is provided by atomic ensembles collectively coupled to off-resonant light via a Faraday interaction.
In this setup, continuous polarimetric measurement of the outgoing light realizes the homodyne-type SMEs we have considered; see, e.g., the recent experiment reported in Ref.~\cite{duan2025concurrent}.
Another possible avenue for implementing such QND interactions is provided by cavity-QED schemes based on continuous measurement of a lossy cavity mode, as proposed in Ref.~\cite{Shankar2019}.

On the other hand, the BTC model is closely related to recent experiments with free-space atomic ensembles showing signatures of phase transitions in driven-dissipative collective-emission dynamics~\cite{Ferioli2023}.
While those experiments were not designed to demonstrate the conditional dynamics or metrological enhancement analyzed here, the environment is the emitted optical field, so photon counting provides a natural route to the photodetection unraveling.
Indeed, recent measurements of time-resolved intensity correlations in similar free-space superradiant ensembles~\cite{Ferioli2025} further indicate that continuous monitoring of the emitted light is experimentally plausible.

To obtain a detailed experimental proposal, better tools will be needed to analyze the impact of finite detection efficiency and other sources of noise, which are inevitable in experiments.
While some numerical techniques have been recently presented~\cite{Khan2025,Yang2025b}, they are computationally demanding, and it is unlikely that they can be applied to study the large-$N$ regime.
A promising alternative would be to develop numerically efficient upper bounds tailored specifically to continuously monitored systems, providing a tighter constraint than the currently available, yet often overly optimistic, general bounds.

Secondly, while our analysis has focused on specific collective spin models, this continuous-monitoring enhancement is not fundamentally restricted to them. It is an open question as to which dissipative time crystal models are suitable for quantum-enhanced sensing. Because a dissipative time crystal inherently features persistent macroscopic oscillations, achieving a metrological advantage requires two main conditions: (i) the frequency of these persistent oscillations must explicitly depend on the parameter to be estimated, and (ii) the continuous measurement must capture information from an observable that displays these oscillations. Provided these conditions are satisfied, more complex dissipative time crystals, such as the interacting fermionic lattice discussed in~\cite{Alaeian2022}, may represent promising systems for quantum metrology with continuous monitoring.
Finally, investigating the impact of measurement-based feedback on dissipative time crystals represents an exciting frontier for further study.

Note that BTCs are not the only route to realizing a time-crystalline phase. Discrete time crystals provide an alternative mechanism for generating long-lived oscillations in many-body systems, though the underlying physics is fundamentally different. Unlike BTCs, where time-crystalline behavior emerges in the dissipative dynamics of an open quantum system, discrete time crystals arise in effectively closed systems governed by unitary evolution. In the latter case, time-translation symmetry is broken by an external periodic drive. Consequently, while the oscillation frequency in BTCs is set by intrinsic properties of the system and its coupling to the environment, discrete time crystals exhibit a subharmonic response, namely the system oscillates with a period that is an integer multiple of the driving period. Discrete time crystals have also been proposed~\cite{Iemini2024floquet,Yousefjani2025b,yousefjani2026nonlinearity,Yousefjani2025Discrete} and experimentally realized~\cite{moon2026sensing} as platforms for quantum sensing. However,   their intrinsically unitary dynamics raises an open problem whether continuous monitoring, used in our setup,  can be implemented without destroying the discrete time-crystalline order. Addressing this problem remains a topic for future studies.

%




\section{Methods}
%
%
%
\subsection{Stochastic master equation describing continuous monitoring by homodyne and photodetection}
\label{a:SME}
We start from a Markovian master equation in the Lindblad form
\begin{align}
        \frac{d\rho}{dt} = -i[\hat{H}, \rho] + \mathcal{D}[\hat{c}]\rho\,.
    \label{eq:Lindblad_ME_SM}
\end{align}
We briefly present here the stochastic master equations corresponding to two paradigmatic examples of unravellings of such master equations; that is, continuous photodetection and continuous homodyne detection (for more details on these equations and in general on continuously monitored quantum systems, we refer to~\cite{AlbarelliPLA2024}). 
For photodetection, the corresponding stochastic master equation (SME) for the conditional state, assuming that the measurement has efficiency $\eta$ reads
\begin{multline}
   d \rho^{(c)} = -i \left[ \hat H, \rho^{(c)} \right] dt -  \frac{\eta}{2} \mathcal{H}[\hat c^\dag \hat c] \rho^{(c)} dt \\ 
 + \biggl( \frac{ \hat c \rho^{(c)} \hat c^\dag }{\Tr \left[ \hat c \rho^{(c)} \hat c^\dag \right]} - \rho^{(c)} \biggr) d N_t + (1-\eta)  \mathcal{D}[\hat c]\rho^{(c)} dt \,,
 \label{eq:SME_PD}
\end{multline}  
where $dN_t$ denotes a Poisson increment with average value $\mathbbm{E}\left[ dN_t \right] = \eta \hbox{Tr}\left[ \rho^{(c)} \right] dt$.

Considering instead continuous homodyne detection characterized by a homodyne phase $\theta$, the corresponding SME reads:
\begin{multline}
d \rho^{(c)} (t) =  -i \left[ \hat H, \rho^{(c)}(t) \right] dt +  \mathcal{D}\left[ \hat c \right] \rho^{(c)}(t) dt  \\
 + \sqrt{\eta} \mathcal{H}[ \hat c e^{i\theta}] \rho^{(c)}(t) dw_t,
  \label{eq:SME_HD}
\end{multline}
corresponding to a measured continuous photo-current $I(t)\,dt = \sqrt{\eta\kappa} \langle \hat{c} e^{i\theta} + \hat c^\dag e^{i\theta} \rangle_t \,dt + dw_t$.
In the equation above, $dw_t$ denotes a Wiener increment, and we have employed the non-linear superoperator $\mathcal{H}\left[ \hat c \right] \rho = c \rho + \rho c^\dag - \hbox{Tr}[\rho (c+c^\dag)]\rho$.

In the case of the BTC dynamics described by Eq.~\eqref{eq:btc_master_equation}, one can derive the corresponding stochastic master equations by substituting $\hat{H} = \omega \J_x$ and $\hat{c} = \sqrt{\frac{2\kappa}{N}} \J_-$. It is then clear that, in the case of homodyne, one should fix $\theta = \pi/2$ in order to obtain a photocurrent $I(t)\propto \langle \J_y \rangle$, and thus bearing information on the oscillations displayed in the time crystal phase.\\
\subsection{Vectorization and diagonalization of the Lindbladian in the extreme time-crystal phase}
\label{a:Lindblad_diagonalization}
We here give the basic definitions and methods that were employed in~\cite{nemeth2025solving} to diagonalize the Lindbladian in Eq.~\eqref{eq:btc_master_equation} in the extreme time-crystal phase ($\omega/\kappa \to \infty$). By defining the superspin operators as 
\begin{equation}
    \hat{S}_\alpha = \J_\alpha \otimes \hat{\mathbb{I}} - \hat{\mathbb{I}} \otimes \J_\alpha^T\,
\end{equation}
with $\alpha=\{x,y,z\}$ and the total superspin operator as $\hat{S}^2 = \hat{S}_x^2+\hat{S}_y^2+\hat{S}_z^2$, one can in fact prove that to first order in $\kappa/\omega$ the 
linearized Lindbladian superoperator can be written as~\cite{nemeth2025solving}
\begin{align}
\mathcal{L} \approx i \omega \hat{S}_x - \frac{\kappa}{2N} (\hat{S}_x^2 + \hat{S}^2) \,.
\label{eq:approx_Lindblad}
\end{align}
As a consequence, the eigenvectors of $\mathcal{L}$ are common eigenvectors of $\hat{S}_x$ and $\hat{S}^2$, and the eigenvalues can be readily calculated, obtaining 
\begin{align}
\lambda_{s,s_x} = -\frac{\kappa}{2N} (s (s+1) + s_x^2) + i \omega s_x \,.
\label{eq:eigenvalues}
\end{align}

We now prove that in this limit, the vectorized operator $|\hat{J}_x\rrangle$ is a right (and left) eigenvector of the Lindbladian. 
First, we show that $|\J_x\rrangle$ is a right eigenvector of $\hat{S}_x$ with eigenvalue $s_x = 0$.
\begin{equation}
    \hat{S}_x|J_x\rrangle = |[\J_x,\J_x]\rrangle = 0
\end{equation}
Next, we demonstrate that $|\J_x\rrangle$ is also a right eigenvector of $\hat{S}^2$
\begin{align}
    \hat{S}^2|\J_x\rrangle &= |[\J_y,[\J_y,\J_x]]\rrangle + |[\J_z,[\J_z,\J_x]]\rrangle \nonumber \\
    &= 2 |\J_x\rrangle
\end{align}
This follows from the commutation relations $[\J_\alpha,\J_\beta] = i \epsilon_{\alpha\beta\gamma} \J_\gamma$. The eigenvalues $\hat{S}^2$ are $s(s+1)$ proving that $|\J_x\rrangle$ is a right eigenvector of $\hat{S}^2$ with $s=1$. It follows that $|\J_x\rrangle$ is also a right eigenvector of $\mathcal{L}^\dag$.

We have thus that $|\hat{R}_{1,0}\rrangle = \mathcal{K}|\J_x\rrangle$ where $\mathcal{K}$ is an arbitrary constant. Due to the bi-orthogonality property of right and left eigenvectors, $\llangle \hat{L}_{s,s_x}|\J_x\rrangle = \hbox{Tr}[\hat{L}_{s,s_x}^\dagger \J_x] = \delta_{s,1}\delta_{s_x,0}/\mathcal{K}$.

We note that Eq.~\eqref{eq:approx_Lindblad} can be written in the standard Lindblad form with a Hamiltonian $H = \omega \J_x$ and Lindblad jump operators $L = \{\sqrt{2 \kappa / N} J_x, \sqrt{\kappa / N} J_y, \sqrt{\kappa / N} J_z$. This means it is a Lindbladian of the form in Eq.~\eqref{eq:generic_Lindblad} with $\kappa_{\text{eff}} = \kappa/N$.

\subsection{Static-limit global QFI of the BTC model}
\label{a:static_limit}
In the static limit $\omega \to 0$, the BTC dynamics reduces to collective
dissipation. We restrict to the symmetric Dicke sector, where all states have
fixed total angular momentum $j=N/2$, so that
$\hat{J}^2=j(j+1)=N(N+2)/4$. The unique steady state is the fully polarized
state
\begin{equation}
    \rho_{\rm ss}=\ketbra{0}{0},\qquad
    \ket{0}:=\ket{j=N/2,m=-N/2}.
\end{equation}
We also define the one-excitation Dicke state $\ket{1}:=\hat J_+\ket{0}/\sqrt N$. To evaluate $f_{\text{global}}$ we start by calcuating
\begin{equation}
    \{\hat{J}_x, \rho_{ss}\} = \frac{\sqrt{N}}{2}(\ketbra{1}{0}+\ketbra{0}{1})
\end{equation}
We find that this is a right eigenoperator of the $\omega = 0$ Lindbladian with eigenvalue $-\kappa$, therefore,
\begin{align}
    \Tr[\hat{J}_x e^{\mathcal{L}\tau}\{\hat{J}_x, \rho_{ss}\}] &= e^{-\kappa \tau}\Tr[\hat{J}_x\{\hat{J}_x, \rho_{ss}\}] \nonumber \\
    &= e^{-\kappa \tau} \frac{N}{2}.
\end{align}
Finally, integrating this, we derive the steady-state global QFI rate in the static limit
\begin{equation}
    f_{\text{global}} = 4\int_0^\infty d\tau e^{-\kappa \tau} \frac{N}{2} = \frac{2N}{\kappa}.
    \label{eq:static_limit}
\end{equation}
This scales linearly with the system size, in contrast with the cubic scaling obtained deep in the time-crystal phase.

\subsection{Relationship between thermodynamical-limit rescaling of the master equation and quantum-enhanced estimation precision}
\label{a:thermodynamiclimit_scaling}
In the master equation for the BTC model we have studied in this paper,
\begin{align}
        \frac{d\rho}{dt} = -i\omega[\J_x, \rho] + \frac{2\kappa}{N}\mathcal{D}[\J_-]\rho\,,
    \label{eq:btc_master_equation_2}
\end{align}
the dissipation rate is scaled by $1/N$, which ensures a well-defined thermodynamic limit ($N\to\infty$) where the dynamics become independent of the system size~\cite{Benatti2016,Benatti_2018,Carollo_2024}. 
If we now perform a time rescaling $\widetilde{t} = t/N$, the new master equation reads 
\begin{align}
        \frac{d\rho}{d\widetilde{t}} &= -i N \omega[\J_x, \rho] + 2\kappa\mathcal{D}[\J_-]\rho\,\nonumber \\
        &= -i \widetilde\omega [\J_x, \rho] + 2\kappa\mathcal{D}[\J_-]\rho\,,
    \label{eq:btc_master_equation_NoTL}
\end{align}
where we have defined a new frequency parameter $\widetilde\omega = N\omega$.
The master equation above has been considered in other works, focusing on the estimation of the parameter $\widetilde\omega$~\cite{cabotContinuousSensingParameter2024}.
In this case, one can easily relate the (quantum) Fisher information for the two parameters obtaining: $F_Q(\omega) = N^2 F_Q(\widetilde\omega)$ (we have slightly changed the notation to make it more explicit the dependence on the parameters). Since the time has also been rescaled, the corresponding Fisher information rates (both the global one and the one related to the signal only) will be related to each other by the equation
\begin{align}
f(\omega) &= \frac{F_Q(\omega)}{T} 
          = \frac{N^2 F_Q(\widetilde\omega)}{N\widetilde{T}}\,  
          = N f(\widetilde{\omega}) \,.
\end{align}
For this reason, the QFI rate obtained from the master equation~\eqref{eq:btc_master_equation_2} acquires an extra factor of $N$ compared to that derived from Eq.~\eqref{eq:btc_master_equation_NoTL}. This is consistent with the $N^{2}$ scaling for the Fisher information rate reported in Ref.~\cite{cabotContinuousSensingParameter2024} for the time-crystal phase—corresponding, in our notation, to the regime $\widetilde{\omega} > N\kappa$—as opposed to the $N^{3}$ scaling derived here.

The fixed environmental scaling was chosen to allow a closer connection to experimental setups where additional spins can be added without changing the system parameters~\cite{Ferioli2023}. However, as stated above, to remain in the time-crystal regime, $\omega$ must increase with $N$.
At a certain point, this driving frequency will become unfeasibly large.
Additionally, the driving frequency is the unknown quantity that we are trying to estimate. It is for these reasons that we have chosen to keep $\omega$ fixed and rescale the environmental coupling. It is also worth mentioning that, if one considers a fixed dissipative coupling $\kappa$, the corresponding Liouvillian gap is not closing in the large $N$ limit, and thus the oscillations always have a fixed decay time; for this reason, the corresponding dynamics does not rigorously correspond to a time crystal.

As a final remark, we note a related—but contrasting—phenomenon discussed recently in the quantum-battery literature. For $N$ two-level atoms collectively coupled to a bosonic mode via a Dicke Hamiltonian with a fixed ($N$-independent) coupling, one observes a collective advantage in the charging power compared to independent charging of the $N$ atoms~\cite{Ferraro2018}. However, if the Dicke coupling is rescaled to ensure a well-defined thermodynamic limit, this advantage disappears~\cite{Farre2020,CampaioliRMP2024}.
In contrast, in our estimation protocol, taking the proper thermodynamic limit not only preserves but even improves the exponent in the super-classical scaling.
\subsection{General dephasing precision}
\label{a:gen_dephasing}
Here, we generalize the analysis in Sec.~\ref{sec:enhancement_generality} to the general case described in Eq.~\eqref{eq:generic_Lindblad}. For concreteness and without loss of generality, we will fix $\vec h$ in the $x$ direction. We start by evaluating the evolution of $\hat{\vec{J}}$ in the Heisenberg picture
\begin{equation}
\frac{d\hat{\vec{J}}}{dt} = i\omega[\J_x, \hat{\vec{J}}] - \frac{1}{2}\sum_i[\J_{\vec{\ell}_i}, [\J_{\vec{\ell}_i}, \hat{\vec{J}}]]
\label{eq:Heisenberg-Lindblad}
\end{equation}
$\{\vec{\ell_i}\}$ are real vectors with components $\{\ell_{i,x},\ell_{i,y},\ell_{i,z}\}$.
We now calculate the unitary and dissipative contributions for each angular momentum operator $\hat{J}_k$, using the standard commutation relations $[\hat{J}_i, \hat{J}_j] = i\epsilon_{ijk}\hat{J}_k$.

The unitary part of the evolution is given by $i[H, \hat{J}_k]$.
\begin{align}
    i[\omega \hat{J}_x, \hat{J}_x] &= 0 \label{eq:Jx_unitary}  \\
    i[\omega \hat{J}_x, \hat{J}_y] &= -\omega \hat{J}_z \label{eq:Jy_unitary}  \\
    i[\omega \hat{J}_x, \hat{J}_z] &= \omega \hat{J}_y \label{eq:Jz_unitary}
\end{align}

We will calculate the evolution for a single jump operator; the total evolution is given by the sum of the individual evolutions. We first compute the inner commutators $[\J_{\vec{\ell}_i},\J_k]$.
\begin{align}
    [\J_{\vec{\ell}_i},\J_x] &=  i\ell_{i,z} \hat{J}_y - i\ell_{i,y} \hat{J}_z \label{eq:comm_JxL}  \\
    [\J_{\vec{\ell}_i},\J_y] &=  i\ell_{i,x} \hat{J}_z - i\ell_{i,z} \hat{J}_x \label{eq:comm_JyL}  \\
    [\J_{\vec{\ell}_i},\J_z] &=  i\ell_{i,y} \hat{J}_x - i\ell_{i,x} \hat{J}_y \label{eq:comm_JzL} 
\end{align}
Now, we use these results to compute the full double-commutator for each operator.
\begin{align}
    [\J_{\vec{\ell}_i}, [\J_{\vec{\ell}_i}, \hat{J}_x]] &= (\ell_{i,y}^2 + \ell_{i,z}^2)\hat{J}_x - \ell_{i,x}\ell_{i,y}\hat{J}_y - \ell_{i,x}\ell_{i,z}\hat{J}_z  \label{eq:dcomm_Jx} \\
    [\J_{\vec{\ell}_i}, [\J_{\vec{\ell}_i}, \hat{J}_y]] &= (\ell_{i,x}^2 + \ell_{i,z}^2)\hat{J}_y - \ell_{i,y}\ell_{i,x}\hat{J}_x - \ell_{i,y}\ell_{i,z}\hat{J}_z  \label{eq:dcomm_Jy} \\
    [\J_{\vec{\ell}_i}, [\J_{\vec{\ell}_i}, \hat{J}_z]] &= (\ell_{i,x}^2 + \ell_{i,y}^2)\hat{J}_z - \ell_{i,z}\ell_{i,x}\hat{J}_x - \ell_{i,z}\ell_{i,y}\hat{J}_y  \label{eq:dcomm_Jz}
\end{align}
We define the summed components
\begin{align}
    A_{\alpha\beta} &= \frac12\sum_i(\ell_{i,\alpha}^2 + \ell_{i,\beta}^2) \\
    B_{\alpha\beta} &= \frac12\sum_i\ell_{i,\alpha}\ell_{i,\beta}
\end{align}

The complete equations of motion can be written compactly in matrix form. The dynamics are described by $\frac{d}{dt}\hat{\vec{J}} = M \hat{\vec{J}}$, where the evolution matrix $M$ is:
\begin{equation}
    M = \begin{pmatrix}
    -A_{yz} & B_{xy} & B_{xz} \\
    B_{xy} & -A_{xz}  & B_{yz} - \omega \\
    B_{xz} & B_{yz} + \omega & -A_{xy}
    \end{pmatrix}
\label{eq:Matrix_EOM}
\end{equation}
This linear system of operator equations fully describes the open quantum dynamics of the angular momentum operators in the Heisenberg picture. This can be solved numerically, but in the limit $\omega \gg \sum_i\|\vec{\ell}_i\|^2$, we can solve it analytically by moving to the rotating frame and applying the rotating-wave approximation (RWA). We define the rotation matrix
\begin{equation}
    M_0 = \begin{pmatrix}
    0 & 0 & 0 \\
    0 & 0 & -\omega \\
    0 & \omega & 0
    \end{pmatrix}
\end{equation}
and the remainder $V = M - M_0$. In the rotating frame we have $\hat{\vec{J}}_R = e^{-M_0 t} \hat{\vec{J}}$, this vector evolves as
\begin{equation}
    \frac{d}{dt}\hat{\vec{J}}_R = e^{-M_0 t}Ve^{M_0 t}\hat{\vec{J}}_R :=V_R \hat{\vec{J}}_R
\end{equation}
Applying the RWA we average $V_R$ over a period of $e^{-M_0 t}$ to get $\overline{V}_R$. This is a diagonal matrix with entries $\{-\Gamma_x, -\Gamma_y, -\Gamma_z\}$
\begin{align}
    \Gamma_x & = A_{yz} \\
    \Gamma_y = \Gamma_z & = \frac{A_{xy}+A_{xz}}{2}.
\end{align}
Moving back to the lab frame, the averaged evolution is
\begin{align}
    \J_x(t) &= e^{- \Gamma_x} \J_x \\
    \J_y(t) &= e^{- \Gamma_y} \left(\cos(\omega t)\J_y - \sin(\omega t)\J_z\right) \\
    \J_z(t) &= e^{- \Gamma_z} \left(\sin(\omega t)\J_y + \cos(\omega t)\J_z\right)
\end{align}
Shifting back to the generic Hamiltonian direction, we can split every $\vec{\ell}_i= \vec{\ell}_{i,\|} + \vec{\ell}_{i,\perp}$ into the parallel and perpendicular parts. Using this decomposition, we find
\begin{align}
    \Gamma_x &= \frac12\sum_i \|\vec{\ell}_{i,\perp}\|^2 \\
    \Gamma_y = \Gamma_z &= \frac14\sum_i \left(2 \|\vec{\ell}_{i,\|}\|^2 + \|\vec{\ell}_{i,\perp}\|^2\right)
\end{align}
It is now simple to calculate the global QFI by following the same reasoning shown in Sec.~\ref{sec:qfi_rate_scaling}, obtaining
\begin{equation}
    f_{\text{global}} =  \frac{2N(N+2)}{3 \Gamma_x}.
\end{equation}
Notice that the rotating wave terms are $\mathcal{O}(\omega^{-2})$ and can therefore safely be ignored for large values of the driving amplitude.

Additionally, we find that observables perpendicular to $\J_{\vec h}$ oscillate at a frequency $\omega$ and these oscillations persist in the thermodynamic limit when $\Gamma_y \to 0$. 

\subsection{Derivation of the ultimate bound for inefficient detection}
\label{a:inefficient_bound}
As we described in the main text and as it is now clear from the form of the SMEs~\eqref{eq:SME_PD} and~\eqref{eq:SME_HD} reported above, in the case of inefficient detection ($\eta<1$), the evolution of the conditional state can be written as in Eq.~\eqref{eq:btc_sme_efficiency}.
As a consequence, one can interpret the unravelling with efficiency $\eta$ as one of the possible metrological strategies aided by auxiliary systems, in the presence of an unavoidable Markovian noise; for the BTC model, it is thus described by the master equation
\begin{align}
\frac{d \rho}{dt} =  -i \omega \left[ \J_x, \rho \right] dt +  \frac{2(1-\eta)\kappa}{N} \mathcal{D}\left[ \J_- \right] \rho\, dt \,.
\label{eq:BTC_ME_eta}
\end{align}
The QFI corresponding to all possible estimation protocols, that encompass measuring the main system itself, measuring auxiliary systems coupled to the main one (in our case, these correspond to the portion of the environment that can be monitored, since $\eta < 1$), as well as arbitrary quantum controls and feedback, can be upper bounded with the techniques introduced in Refs.~\cite{demkowicz2017adaptive,Wan2022,Kurdzialek2023a,Das2025a}.
Such QFI bounds are tight at long times, since they are attainable with suitable error correction strategies~\cite{Zhou2018,Zhou2019e}.
While our specific protocol based on continuous monitoring is not guaranteed to be optimal and saturate these bounds, they are useful to understand the fundamental limitations imposed by inefficient detection.

In particular, the Lindblad master equation~\eqref{eq:BTC_ME_eta} corresponds to a situation in which, regardless of the protocol applied, a quadratic scaling in time of the QFI is forbidden at long times.
This happens because the parameter-encoding Hamiltonian $\hat{J}_x$ is ``too similar'' to the noise $\hat{J}_{-}$ and so it cannot be corrected to restore a long-time quadratic scaling.
More formally, the relevant mathematical conditions is that the parametric derivative of the Hamiltonian, $\hat{J}_x$, belongs to the so-called Lindblad span: $\mathrm{span}_{\mathbb{C}}\left\{ \hat{\mathbb{I}} , \hat{J}_{-},  \hat{J}_{+} ,  \hat{J}_{+} \hat{J}_{-} \right\}$.

When this condition is met, the following bound can be derived~\cite{demkowicz2017adaptive,Wan2022,Das2025a}:
\begin{align}
F_Q(\omega) \leq 4 T \min_{\gamma_1,\gamma_2, \gamma_3} \lVert \hat{\alpha} \rVert , \quad \text{s.t. } \hat{\beta} = 0
\end{align}
where $\lVert \hat{A} \rVert$ denotes the operator norm, and the operator $\hat{\alpha}$ reads
\begin{align}
\hat{\alpha} &= |\gamma_1|^2 \,\hat{\mathbb{I}} + \gamma_2 \sqrt{\frac{2(1-\eta)\kappa}{N}} ( \gamma_1^* \hat{J}_+ + \gamma_1 \hat{J}_-) \nonumber \\
&\,\,\,\,\,\, + \gamma_2^2 \frac{2(1-\eta)\kappa}{N} \hat{J}_+ \hat{J}_- \,,
\end{align}
while the operator $\hat{\beta}$ defining the linear constraint is
\begin{align}
    \hat{\beta} &= \sqrt{\frac{2(1-\eta)\kappa}{N}} (\gamma_1^* \hat{J}_+ + \gamma_1 \hat{J}_-) + \gamma_2 \frac{2(1-\eta)\kappa}{N} \hat{J}_+ \hat{J}_- 
    \nonumber \\
    &\,\,\,\,\,\,\, + \hat{J}_x + \gamma_3 \hat{\mathbb{I}}  \, . \label{eq:beta}
\end{align}
These operators depend on the free parameters $\gamma_1 \in \mathbbm{C}$ and $\gamma_2, \gamma_3 \in \mathbbm{R}$, to be optimized.
In order to have Eq.~\eqref{eq:beta} equal to zero (notice that this can be satisfied only if $\hat{J}_x$ belongs to the Lindblad span), it is straightforward to observe that one has to fix $\gamma_2 = \gamma_3 = 0$ and 
\begin{equation*}
\gamma_1 = - \frac{1}{2} \sqrt{\frac{N}{2(1-\eta)\kappa}} \,.
\end{equation*}
As a consequence, one has that $\lVert \hat{\alpha} \rVert = N / (8(1-\eta)\kappa)$ needs no further optimization, and the upper bound on the ratio between the QFI $F_Q(\omega)$ and the evolution time $T$ reads
\begin{align}
\frac{F_Q(\omega)}{T} \leq \frac{N}{2(1-\eta)\kappa} \,. \label{eq:DDbound}
\end{align}
Notice that, as we mentioned above, this bound applies to estimation strategies involving measurement not only on the auxiliary systems (in our case, the environment monitored with efficiency $\eta$), but also on the system itself.
Therefore, the bound above also applies when one considers the environment monitoring as the only source of information on the parameter $\omega$, and thus one can write
\begin{align}
f_{\textrm{signal}} \leq \frac{N}{2(1-\eta)\kappa} \,. \label{eq:bound_eta_signal}
\end{align}
As already mentioned, in this general scenario, the bound is attainable for large $T$ through approximate error correction~\cite{Zhou2019e}, which reduces the noise to an effective dephasing.
Necessary and sufficient conditions for saturating this fundamental bound only with continuous monitoring and without direct access to the system are not known.
However, at least one example where the fundamental bound is saturated is known~\cite[Sec.~III.A and Appendix~D]{Gorecki2025a}.

Regarding the TCD model, one could follow the same line of reasoning and consider the master equation
\begin{align}
\frac{d \rho}{dt} =  -i \omega \left[ \J_x, \rho \right] dt +  \frac{2(1-\eta)\kappa}{N} \mathcal{D}\left[ \J_z \right] \rho\, dt \,.
\label{eq:TCD_ME_eta}
\end{align}
In contrast with from the BTC model, for the TCD model the operator $\hat{J}_x$ that encodes the parameter does not belong the Lindblad span, which this time reads $\mathrm{span}_{\mathbb{C}}\left\{ \hat{\mathbb{I}} , \hat{J}_{z}, \hat{J}_{z}^2 \right\}$.
In this situation, the long-time upper bound is quadratic in $T$ and can be computed similarly by minimizing an operator norm.
More specifically, for the master equation in Eq.~\eqref{eq:TCD_ME_eta} the relevant bound is computed as~\cite{Das2025a}
\begin{equation}
   F_Q(\omega) \leq  T^2 4 \left( \min_{\gamma \in \mathbb{R}} \left \Vert \hat{J}_x + \frac{2 \kappa (1-\eta) }{N} \gamma \hat{J}_z^2 \right \Vert \right)^2 = N^2 T^2
\end{equation}
which is achieved for $\gamma = 0$, for which $\Vert \hat{J}_x \Vert = \frac{N}{2}$.
We see that for this model the fundamental upper bound matches the noiseless case; thus, in principle, the effect of the noise can be completely counteracted by error correction.
While this result does not imply any useful new constraint for protocols based on continuous monitoring, where typically the QFI scales linearly with $T$, crucially it does not forbid a superlinear scaling in $N$, unlike for the BTC.
%
\section*{Data availability}
The datasets generated and analyzed during the current study are not publicly available as they have not yet been organized for public archiving, but are available from the corresponding author on reasonable request.
\section*{Code availability}
Code used to generate data in this study are available from the corresponding author upon reasonable request.
%
\bibliographystyle{sn-nature}
\bibliography{refsNature}
\section*{Acknowledgments}
EOC, MGAP and MGG acknowledge support from MUR and Next Generation EU via the PRIN 2022 Project CONTRABASS (Contract N. 2022KB2JJM), NQSTI-Spoke2-BaC project QMORE (contract N. PE00000023-QMORE), NQSTI-Spoke1-BaC project QSynKrono (contract n. PE00000023-QuSynKrono). MGG acknowledges support from the Ministry of Science and Technology of China via the {\it Talented Young Scientist Program}. MGAP acknowledges support from the “111 Project” (No. B21045).
FA acknowledges financial support from Marie Skłodowska-Curie Action EUHORIZON-MSCA-2021PF-01 (project \mbox{QECANM}, grant n. 101068347).
VM acknowledges partial funding by Khalifa University of Science and Technology through the Project ID: KU-INT-RIG-2024-8474000739 and thanks support from the National Natural Science Foundation of China Grants No. 12374482 and No. W2432005. AB acknowledges support from National Natural Science Foundation of
China (Grants No. 12050410253, 92065115, and 12274059).
\section*{Author contribution}
M.G.G. conceived and supervised the project. E.O.C. performed analytical and numerical calculations. V.M. performed numerical calculations. F.A., M.G.A.P and A.B. provided additional supervision. All authors discussed the results, contributing to their interpretation and presentation. E.O.C. wrote the first draft of the manuscript. All authors contributed to reviewing and editing the manuscript and approved the final version.
\section*{Competing Interests} 
Author A.B. is Associate Editor of npj Quantum Information. A.B. was not involved in the journal’s review of, or decisions related to, this manuscript.
The other authors declare no competing interests.

\onecolumngrid
\clearpage
\appendix
\section*{Supplementary Material}
\setcounter{equation}{0}
\renewcommand{\theequation}{S\arabic{equation}}
\setcounter{figure}{0}
\renewcommand{\thefigure}{S\arabic{figure}}
\subsection*{Supplementary Note 1: Global QFI rate of the BTC model at the critical point}
\label{a:globalQFI_critical}

\begin{figure}[t]
\begin{center}
\includegraphics[angle=0,width=0.98\linewidth]{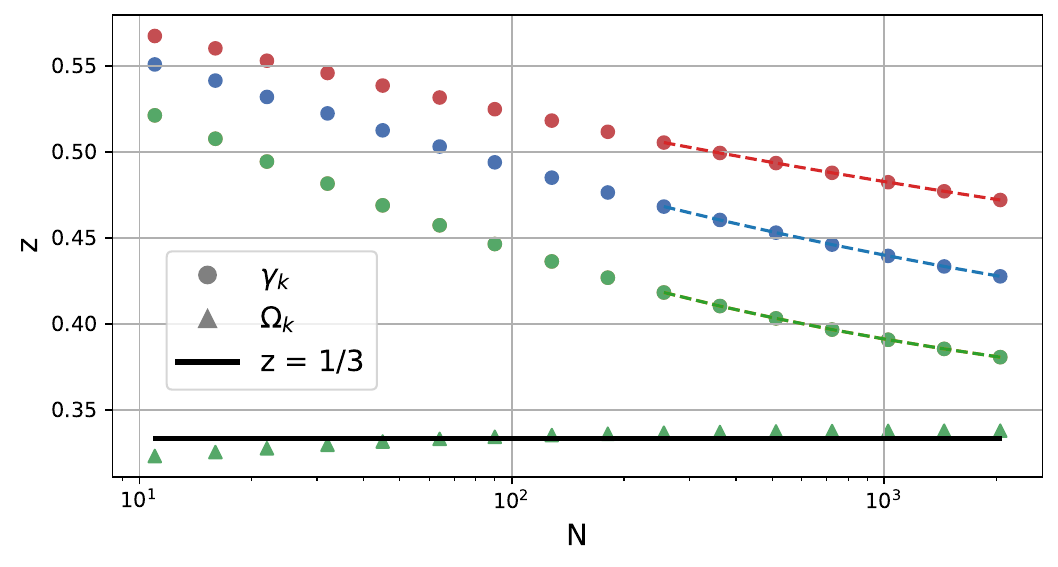} 
\end{center}
\caption{The scaling exponent $z_{k,N}$ of the three smallest eigenvalues with non-zero contribution to $f_{\text{global}}$ as a function of $N$ (circles). The equivalent scaling exponent of the imaginary parts of the eigenvalues is also shown when the eigenvalues have non-zero real part (triangles). The dashed lines show a power law scaling fit to $z_{k,N}$ for each eigenvalue. The black solid line shows $z = 1/3$, the value of the limit of all the power law scaling fits.}
\label{f:z_scaling}
\end{figure}

At the critical point $\kappa/\omega = 1$, so the perturbative approach used for the time-crystal phase is no longer valid. However, the behaviour of the smallest eigenmodes is still informative. In analogy with the approach taken in the time-crystal phase, we can diagonalize the Lindbladian as $\mathcal{L}=\sum_k\lambda_k |\hat{R}_k \rrangle \llangle \hat{L}_k|$ and express the correlation function in the form
\begin{align}
    f_{global} &= 8 \Re \sum_{k, \lambda_k \neq 0} \frac{1}{\lambda_k} \llangle \delta \hat{O} | r_k \rrangle \llangle l_k | \delta \hat{O} \rho_{\text{ss}} \rrangle. \nonumber \\
    &:= 8 \sum_{k, \lambda_k \neq 0} \Re \frac{A_k}{\lambda_k}
    \label{eq:fg_vector}
\end{align}
with $\lambda_k = -\gamma_k + i \Omega_k$. In the large $N$ limit, the amplitudes and eigenvalues will have fixed scaling behaviour with $N$; we label these $|A_k| \sim N^{-2\Delta_k}$, $\gamma_k\sim N^{-z_k}$ and $\Omega_k \sim N^{-Z_k}$. By analyzing the scaling of the eigenvalues with smallest real part and the corresponding amplitudes, we can estimate their contribution to the global QFI. These critical exponents do not converge to a single value in the range of $N$ we can currently simulate. However, we can make extrapolations based on the finite scaling analysis. Since we can calculate $|A_{k,N}|$ and $\lambda_{k,N}$ with high numerical precision we would expect each additional neighboring pair of $|A_{k,N}|$ and $|A_{k,N^\prime}|$ to provide a more accurate estimate of the large $N$ scaling. We therefore define
\begin{align}
    z_{k,N} &= \frac{\log(\gamma_{k,N}/\gamma_{k,N^\prime})}{\log(N/N^\prime)} \label{eq:z_scaling}\\
    \Delta_{k,N} &= \frac12\frac{\log(|A_{k,N}|/|A_{k,N^\prime}|)}{\log(N/N^\prime)} \label{eq:A_scaling}
\end{align}
Figure~\ref{f:z_scaling} shows $z_{k,N}$ for the 3 smallest eigenvalues with non-zero $A_k$. It is clear that the critical exponents are not close to converging to a fixed value; however, we can fit $z_{k,N}$ with a power law decay, $z_{k,N} = a_k + b_kN^{-x_k}$ for large values of $N$. We find that the best fit is given by $a_k \approx 1/3$ for all three $\gamma_k$. Additional evidence is provided by the scaling of $\Omega_k$ for the smallest complex eigenvalue, which converges much faster. By a similar analysis, we find that $\Delta_k \approx -2/3$ for these eigenmodes. Combining these two results, we approximate the contribution of these eigenmodes to the global QFI scaling to be $f_{\text{global}} \sim N^{5/3}$

\begin{figure}[t]
\begin{center}
\includegraphics[angle=0,width=0.98\linewidth]{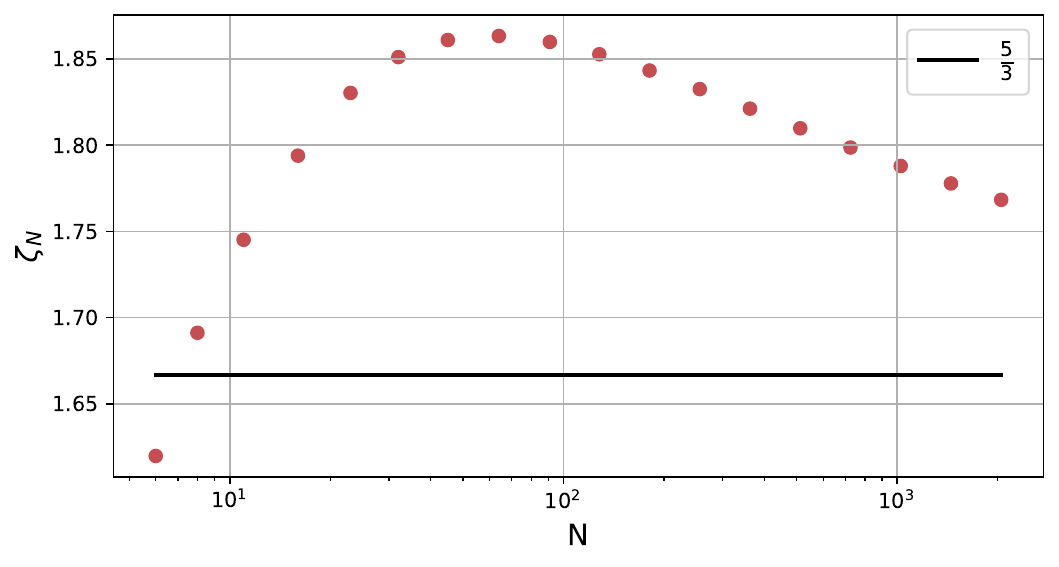} 
\end{center}
\caption{The scaling exponent $\zeta_{N}$ of $f_{\text{global}}$ as a function of $N$ (circles). The black solid line is at $\zeta = 5/3$, a lower bound and proposed estimate of the value of $\zeta_N$ in the large N limit.}
\label{f:zeta_scaling}
\end{figure}

Unfortunately, we do not have enough numerical power to perform a power law fit directly on the global QFI rate; however, we can still learn something from the scaling. In analogy to Eqs.~\eqref{eq:z_scaling} and~\eqref{eq:A_scaling} we define $\zeta_N$ to capture how the scaling exponent of $f_{\text{global}} \sim N^{\zeta_N}$ scales with $N$. In Fig.~\ref{f:zeta_scaling} we see that $\zeta_N$ peaks around $1.87$ then begins to decline. We propose that the most likely final value in the large $N$ limit is $\zeta = 5/3$.

For generic quantum critical points the correlation function, $C(t) =\langle\{\delta\hat{O}(t), \delta\hat{O}(0)\}\rangle \rho_{ss}]$, takes a universal form~\cite{Ilias2022}:
\begin{equation}
    C(t) = L^{d-2\Delta_{\hat{o}}}\phi_L(L^z t^{-1}).
\end{equation}
Here, $d$ is the dimension of the lattice, and $L$ is the size of each of these dimensions. $z$ is the scaling exponent of the Liouvillian gap. The BTC is a system of $N$ non-interacting spins and therefore has lattice dimension 0 and $N$ takes the role of ``system size'', so we expect $C(t) = N^{-2\Delta_{\hat{O}}}\phi_N(N^z t^{-1})$. This results in an overall QFI scaling
\begin{equation}
    f_{\text{global}} \sim N^{z -2\Delta_{\hat{O}}}
\end{equation}
However, when we calculate $C(t)$, we find this universal scaling absent, suggesting that the critical point in the BTC is non-generic. 

There are multiple ways to calculate $\Delta_{\hat{O}}$, the most straightforward is to analyze the scaling of $C(0) = \hbox{Tr}\left[\J_x^2 \rho_{ss}\right] $ with $N$. Using this method we find that $\Delta_{\hat{O}} \approx -5/6$. $\Delta_{\hat{O}}$ can also be extracted from the susceptibility of local observables at the critical point~\cite{Ilias2022}, $\partial_\omega \langle\hat{o}\rangle \propto N^{1/\nu - \Delta_{\hat{O}}}$. We know that $\nu = 3/2$~\cite{montenegroQuantumMetrologyBoundary2023} allowing us to extract $\Delta_{\hat{O}} \approx -2/3$ from both $\hat{o} = \J_y$ and $\hat{o} = \J_z$. These results tell us that the final global QFI scaling is most likely between $N^2$ and $N^{5/3}$, with our numerical analysis pointing clearly in the direction of $N^{5/3}$. The multiple values of $\Delta_{\hat{O}}$ highlight the non-generic nature of the critical point and the difficulty in determining the global QFI scaling via critical exponents. 

\end{document}